# MODELING OF BASIC PHYSICS ISSUES IN TOROIDAL PINCHES AND TOOLS FOR PERFORMANCE CONTROL

S. CAPPELLO, D. BONFIGLIO, G. DI GIANNATALE, A. KRYZHANOVSKYY, G. MANDUCHI, A. RIGONI GAROLA, F. SATTIN, L. SPINICCI, G. SPIZZO, M. VERANDA, N. VIVENZI,
RFX Consortium (CNR, ENEA, INFN, Padova University, Acciaierie Venete SpA)
Padova, Italy
Email: susanna.cappello@cnr.it

D.F. ESCANDE
Aix-Marseille Université, CNRS, PIIM, UMR 7345
Marseille, France

L. CHACÓN,
Los Alamos National Laboratory,
Los Alamos, New Mexico, USA

D. GRASSO,
ISC-CNR and Politecnico di Torino, Energy Department,
Torino, Italy

M.V. FALESSI,
ENEA, Fusion and Nuclear Safety Department,
Frascati, Roma, Italy

F. PEGORARO
Pisa University, Physics Department
Pisa, Italy

**Abstract**

Nonlinear modelling has improved significantly in dealing with physics issues encountered in magnetically confined toroidal pinches, such as the Reversed Field Pinch and Tokamak configurations. Here, recent results relevant to the understanding of helical self-organization processes are presented: formation of internal transport barriers, temporary loss of operational point, relaxation-reconnection events, excitation of Alfvén waves, and a possible fundamental mechanism for ion heating in plasmas. Starting from physics processes in Reversed Field Pinch plasmas, several similarities with Tokamaks are presented. In addition, data analysis tools, machine learning ”autoencoding” techniques, are here trained for the first time on a RFP data analysis case.

## 1. INTRODUCTION

In the last few years, nonlinear modelling and data analysis tools have improved in several respects, in particular to deal with physics issues encountered in magnetically confined toroidal pinches, such as the Reversed Field Pinch and Tokamak configurations. Benchmark-verified codes for 3D nonlinear MHD basic modelling find a reasonable comparison (validation) against a number of experimental observations in the RFX-mod and other devices. The realistic description obtained within our basic modelling might provide a useful set of means to train and validate advanced data analysis tools, like machine learning techniques, with the aim of understanding and optimizing magnetic configurations. Most recent progresses are reported in this manuscript, relying on extensive exploitation of the simple 3D nonlinear visco-resistive modelling as implemented in the SpeCyl code [see benchmark paper [1] and references therein], describing current-driven dynamics typical of pinch configurations in cylindrical geometry. Magnetic topology studies are based on the Field Line Tracing code NEMATO [which has been verified against the guiding center code Orbit, see [2] and therein references], and a new refined tool to detect Lagrangian Coherent Structures (see benchmark verification with earlier tool in [3]) is used here for comparison with results from a temperature equation solver [4].

## 2. HELICAL SELF-ORGANIZATION IN 3D NONLINEAR MODELING; ON THE ROLE OF BOUNDARY CONDITIONS.

After highlighting the key role of helical shaping of the magnetic boundary in RFPs [5] and Tokamaks [6], and the discovery of new RFP helical states excited by suitable seed edge magnetic perturbations [7], the implementation of more realistic boundary conditions, including resistive shell and vacuum layer, provided two major additional results relevant for RFPs. First, the decrease of secondary modes is predicted when increasing the shell-plasma proximity, as expected to occur in the upgraded RFX-mod2 device, expected to start operation in 2022 [8]. Then, for given thin shell resistivity values, intermittent helical regimes emerge akin to medium current RFX-mod discharges [9]. Fig. 1 a) and b) summarize the results obtained with helical shaping of magnetic boundary (i.e. by application of the seed Magnetic Perturbation, MP, mimicking the residual average finite value of edge radial magnetic field $b_{r\,(1,-7)}$ observed in RFX-mod discharges). Fig. 1a) shows the scaling with the Hartmann number[1], $H = (\nu\eta)^{-1/2}$, of the dominant-mode's average amplitude in a set of simulations without/with the MP. Note that, according to classical estimates of the perpendicular viscosity, $H$ is expected to increase with plasma current [11]. Black dots (no MP) represent the well-known decrease of mode amplitudes with H (or plasma current). Colored dots show that the MP-boosted mode stops decreasing (while secondary modes do, similarly to the black dots) once reached a saturation amplitude proportional to the MP amplitude. This shows that the MP threshold needed to excite a QSH regime decreases with $H$, i.e. with increasing plasma current, as observed in experiments. Moreover, Fig. 1 b) shows that the temporal fraction during which a RFP plasma finds in a QSH state increases with $H$, another result well confirmed by experiments on the RFX-mod and MST [12, 13]. Fig. 1 c) shows the intermittent (m,n)=(1,-7) regime resulting for a specific choice of vacuum layer (b/a=1.2) and thin shell characteristic time (without externally applied MP).

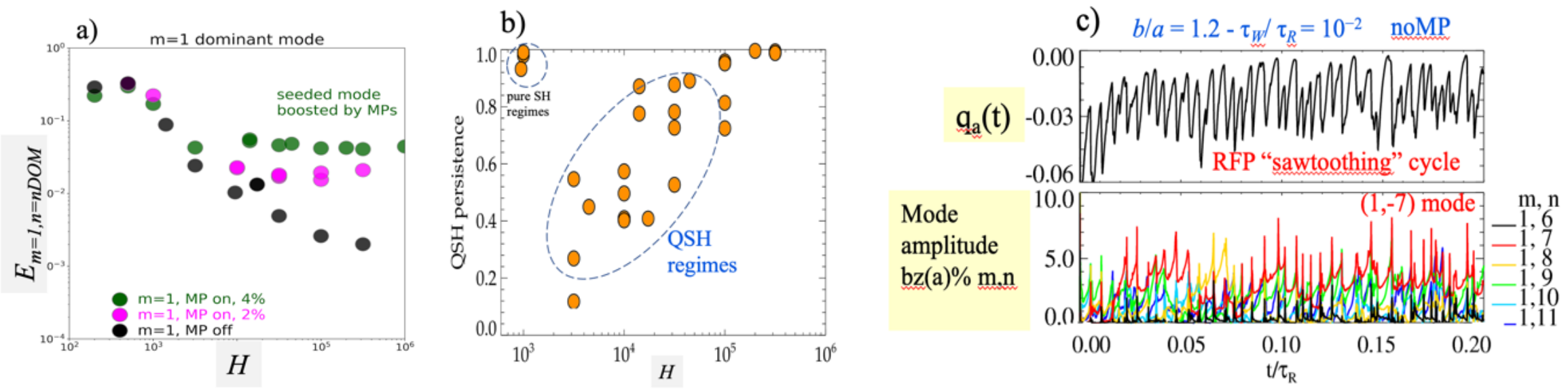

*FIG. 1. Effect of Boundary Conditions on RFP dynamical regimes.* ***a)*** *time averaged amplitude of m=1 dominant mode: black dots represent the natural decrease of mode amplitudes with H (increasing plasma current), coloured dots show that the boosted mode with applied Magnetic Perturbation, MP, stops decreasing once a saturation amplitude proportional to the applied MP amplitude is reached.* ***b)*** *The temporal quasi helical regime incidence in simulations with MP show a clear increasing trend with H, similar to the one observed in RFX-mod experiments in the interval of 0.8-1.2 MA (compare highlighted dots with Figs. 4 and 5 of ref. [12])* ***c)*** *Self-organized intermittent 1/7 helical regime in RFP simulations with resistive thin shell and vacuum layer (no applied MP).*

## 3. LAGRANGIAN COHERENT STRUCTURES (LCS) AND TEMPERATURE GRADIENTS IN CHAOTIC DOMAINS

Refined techniques to detect Lagrangian Coherent Structures (LCS) have been recently developed [14]. LCS rule the "motion" of magnetic field lines and can provide the signature of Cantori structures able to produce barriers to magnetic field lines wandering even within a chaotic domain [15,3,16]. The application to an RFP simulation case is shown here. Fig. 2 shows that the LCS detected in a schematic 3D nonlinear visco-resistive RFP simulation case, whose Poincaré plot can be observed on the left side of the picture, are located close to temperature gradients calculated by a temperature equation solver [4], including anisotropic thermal conductivity. Note that such temperature gradients are located in regions where the magnetic field lines would seem to behave in a stochastic fashion, contrary to naïve expectations about stochastic magnetic fields.

---

[1] with η,ν dimensionless resistivity and viscosity, see ref [10].

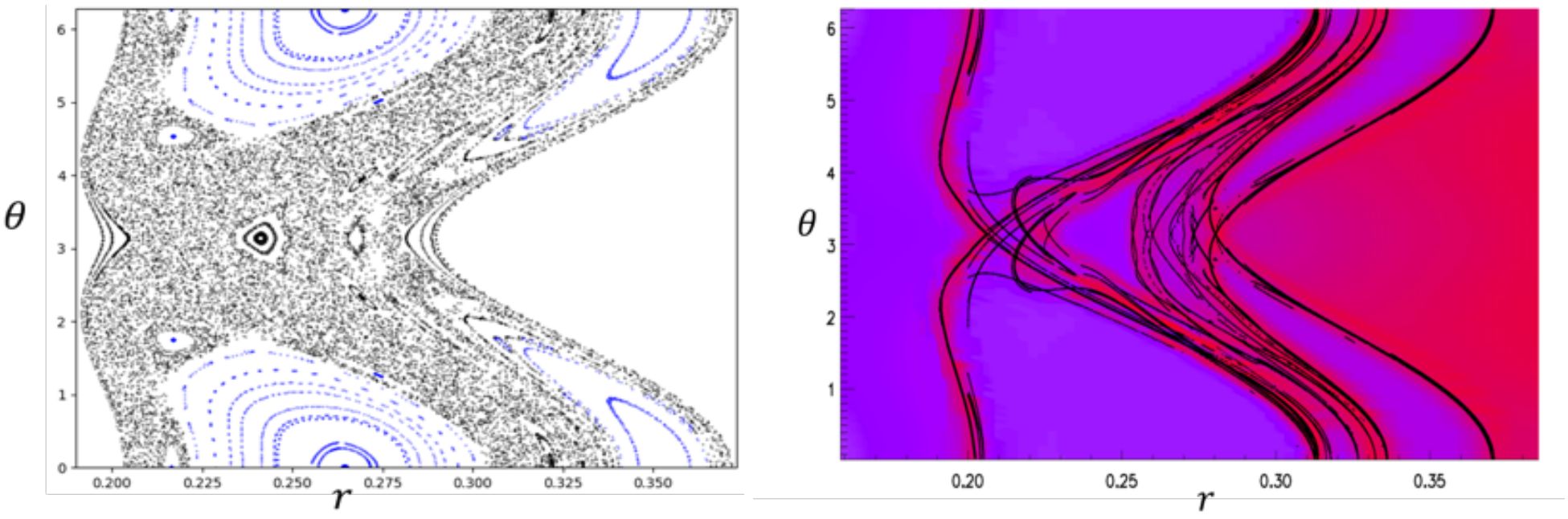

*FIG. 2. (Left) Poincarè surface of section, and (right) Lagrangian Coherent Structures (black curves) and temperature gradient (colour map). The anisotropy ratio employed to obtain the picture on the right is $\chi_{\parallel}/\chi_{\perp} = 10^7$.*

For this reason, LCS detection represent the best candidate to explain the formation of the electron temperature barriers observed in RFX-mod [17]. In view of experimental validation of this advanced diagnostic LCS tool, a data analysis -machine learning- (ML) technique has been used for the first time on the numerical RFP simulation data. In particular, as a first step, the issue of fast and effective reconstruction of the modes' profiles is addressed. Fig. 3 shows an instance of the first results of the algorithm training on the model. In this case, the reconstruction shows the dominant (m,n)=(1,-6) and secondary (m,n)=(1,-8) modes (respectively internally non resonant and resonant modes) during the formation of a realistic QSH regime [11] obtained in 3D nonlinear MHD simulation. The eigenfunctions have been reconstructed back from the ML algorithm using the sole values of the magnetic field at the boundary. The ML algorithm exploited a deep neural network (DNN) and it has been trained in two separate stages: at first, a variational-autoencoder (VAE) has been used to represent all input eigenfunctions into a lower dimensional space (the so-called latent space). In a second step, a simple feedforward Neural Network (NN) was then trained to map the edge value of the magnetic field into the corresponding point of the latent space of VAE [18]. The complete reconstruction proceeds then through a former mapping of the boundary condition into a latent space point and the generation of this point back to an eigenfunction set done by the VAE decoder. A similar approach has been adopted in [19].

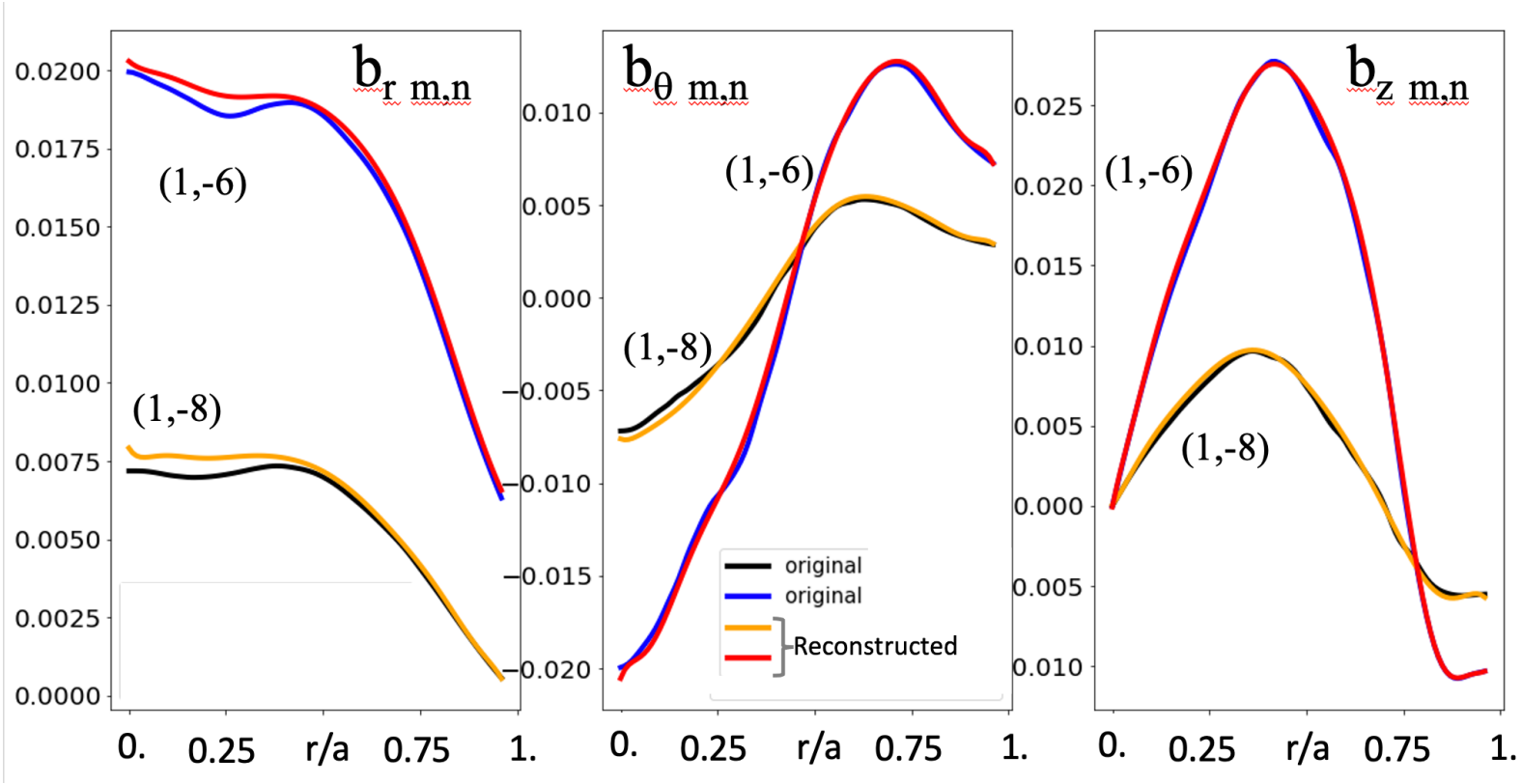

*FIG. 3. Magnetic field components by 3D nonlinear MHD: original in blue and black, reconstruction in red and orange*

## 4. 3D NONLINEAR MHD STUDIES OF RFP AND TOKAMAK SAWTOOTHING.

In both Tokamaks and RFPs, at low dissipation, the transition to quiescent helical regimes ("snakes" in the tokamak case) is somehow preceded by a sawtoothing behaviour, which can be "tuned" or mitigated by applying suitable small edge magnetic field perturbation, MP, at the boundary. The sawtoothing activity is characterized by reconnection events involving magnetic energy conversion into kinetic one, formation of localized structures (intense current sheets, mode phase locking, plasmoids formation [20]) and excitation of Alfvén waves. In particular, the sawtoothing period in the simulations has been shown to depend on the Hartmann number for both tokamaks and RFPs [11], a result to be validated against RFX-mod2 experiments in the next future.

A recent thorough analysis relates the trigger of the cyclic reconnnection event occurring at the RFP sawtooth crash [20] highligthing the role of mode phase locking as crash precursor and the role of a shrinking of the cylindrical safety factor radial profile. Further scrutination of the simulation cases reveals that the helical safety factor related to the helical "quasi-equilibrium" achieved during the QSH cycles, e.g. $(m,n_{QSH})=(1,7)$, tends to flatten in the core and touch the rational value 1/8 prior to the crash. This is shown for a typical simulation case in the left panel of Fig. 4. There, the value of $q_{hel}$ on the helical axis ($\rho = 0$) and the maximum value of $q_{hel}$ are shown in blue and red curves, respectively. The $q_{hel}$ values at $\rho = 0.5$ and $\rho = 1$ (last closed surface at the edge) are shown in black. It is observed that before each crash of the helical configuration (corresponding to times of rapid deepening of the edge $q_{hel}$) both the red and blue curves touch the 1/8 helical resonance. This happens in the shadowed time windows, where the core $q_{hel}$ profile changes from non-monotonic to nearly flat and monotonic. This suggests that the first contiguous mode can trigger the relaxation due its (helical) instability.

Tokamak-like simulations (cylindrical approximation with aspect ratio R/a=4 of a circular Tokamak) shown in the center and right panel of Fig. 4, demonstrate the clear "tuning" effect previously mentioned: a MP with amplitude of 0.05% of the total field at the edge is able to reduce by a half both the period and the excursion of the core-value of the safety factor, as seen comparing the center and right panel of the picture. A similar behaviour can be obtained in RFP also by tuning the helical twist of the applied MPs [11].

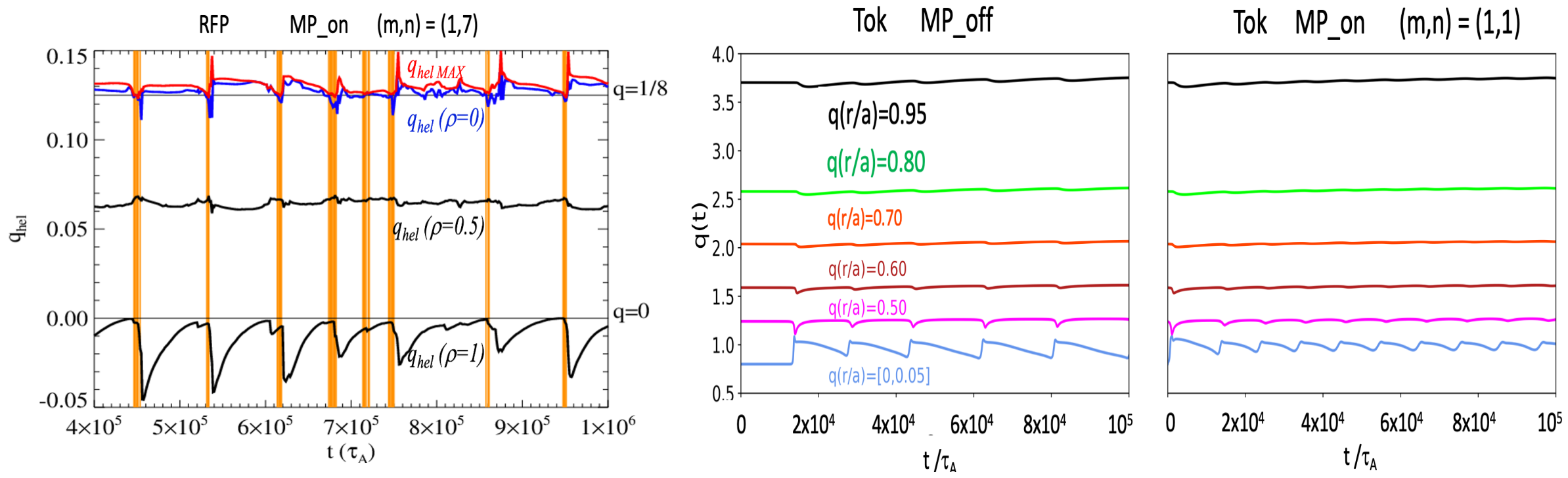

*FIG. 4. Left panel, RFP case: Temporal evolution of the helical safety factor ($q_{hel}$) at different effective radii $\rho$ for a simulation with MP on the 1,-7 mode and $\eta = 10^{-7}, \nu = 10^{-4}$, reported in* [5].
*Center and right panels, TOKAMAK cases: temporal evolution of the safety factor $q(r)$ at different radii r for a simulation without (center) and with MP (right) on the 1,-1 mode, and $\eta = 10^{-6}, \nu = 10^{-4}$.*
*MP amplitude in the RFP (to excite the helical regime) is 2% of the poloidal field at the edge, while in the TOKAMAK case the MP (to mitigate the sawtoothing activity) is 0.8% of the poloidal field at the edge.*

## 5. EXCITATION OF ALFVÈN WAVES

An extensive study of the dynamics of Alfvén waves in nonlinear MHD modelling is also ongoing work for both the RFP and the tokamak-like configuration. An interesting observation is that magnetic reconnection events can excite Alfvén eigenmodes in both the configurations, even in the absence of energetic particles (which are not included in our modelling tool). Typical Alfvénic spectra observed after magnetic reconnection events in the RFP and the tokamak are shown in Fig. 5. The spectrum of the Alfvén eigenmodes excited in RFP configurations is in reasonable quantitative agreement with experimental findings in the RFX-mod device [21]. The modelling supports the interpretation of the two high-frequency modes observed in RFX-mod as a global Alfvén eigenmode (GAE) and the first compressional Alfvén eigenmode (CAE), both with dominant m=1, n=0 character. In the cylindrical circular tokamak case, more recent simulations show that magnetic reconnection events involve plasmoids formation at the q=1 resonant surface and triggers Alfvén eigenmodes (in particular the m=1, n=0 GAE) like in the RFP. Such simulations are similar to those reported in [20] with safety factor equal to $q_0$=0.8 on-axis and $q_a$=4 at the circular boundary. These simulations are characterized by a 3D spectrum of Fourier harmonics instead of a 2D (helical) one, and by a small magnetic perturbation (with perturbed edge radial magnetic field set to $10^{-4}$ of the mean toroidal field) on the marginally stable m=2, n=1 tearing mode to allow for the nonlinear couplings between different helicities. A more complete characterization of these simulations (including investigation of the magnetic field topology) is ongoing.

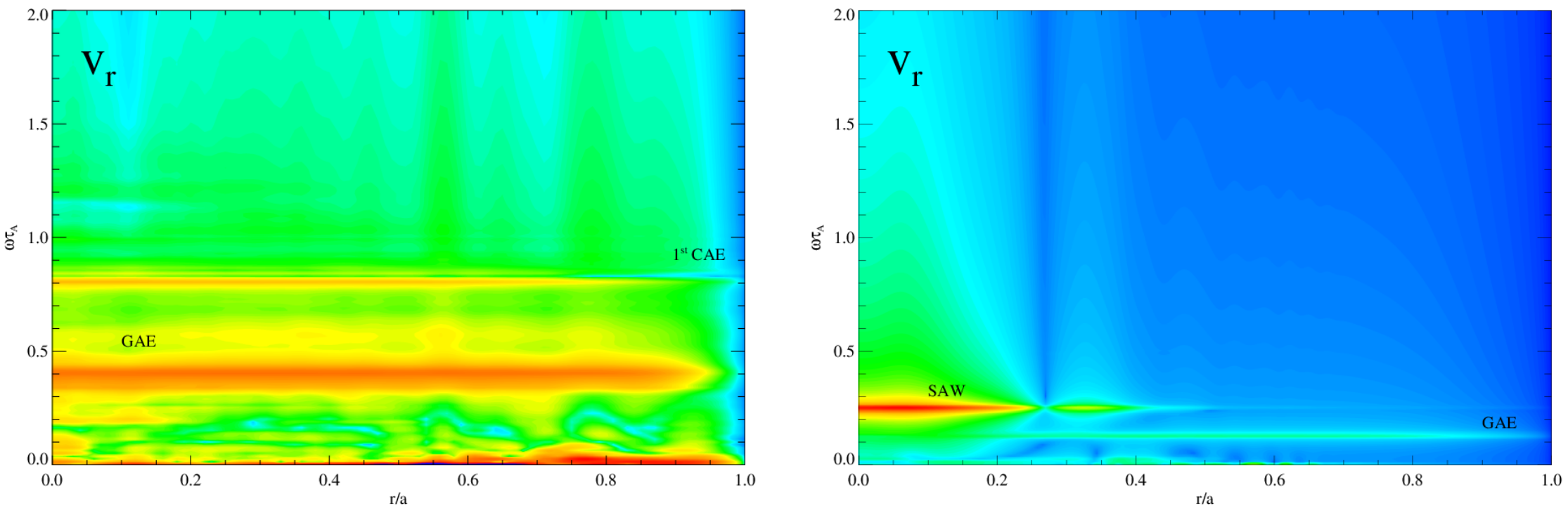

*FIG. 5. Spectrograms of radial velocity perturbations (m=1, n=0 Fourier component) in MHD modelling after a magnetic reconnection event in RFP (left panel) and tokamak configuration (right panel). The colour scale is the same in both panels. Excitation of Alfvén eigenmodes, quite strong in the RFP configuration, is significant in the tokamak configuration as well when a (2,1) mode is present. In particular, triggering of the global Alfvén eigenmode (GAE) is observed in both the configurations. Visco-resistive parameters used in these simulations are as follows: RFP simulation: dimensionless resistivity $\eta = 10^{-6}$ and viscosity $\nu = 10^{-4}$; tokamak simulation: $\eta = 10^{-8}, \nu = 10^{-6}$.*

## 6. STUDIES OF THE ADIABATIC INVARIANTS

Adiabatic invariants (and notably the most important among them, the magnetic moment) are an invaluable tool in particle dynamics studies, at the basis of gyrokinetic formalism. They are associated to the existence of a wide separation between the particle' fast small-scale Larmor rotation and the slower, longer scales of the equilibrium magnetic field. Yet, near future scenarios will more and more often involve fast particles--be they components of high-energy heating beams or fusion products, and will challenge these postulates. An assessment of the validity of the hypotheses which support the existence of adiabatic invariants even in these forthcoming regimes becomes therefore relevant (see, e.g., [22]).

One scenario involving breakup of adiabatic invariants has been known since the early days of fusion research, it involves null points of the magnetic field in cusp fields, where—obviously—the local Larmor radius diverges [23]. Some variants have been afterwards developed, dropping the requirement of a vanishing field, but retaining the basic idea of a locally large ratio $\rho/L$ or $\omega\, T$, with $\rho, \omega$, Larmor radius and frequency, and $L, T,$ the equilibrium local length and time scales. In Hamiltonian dynamics particle trajectories track curves at constant energy $E = H$ in phase -space $(q_i p_i)$. Let $H$ be a smooth Hamiltonian whose phase portrait in the $(q_i p_i)$ space involves one or more separatrices. In correspondence of the hyperbolic X-points any perturbation leads to sensitive dependence upon initial conditions, entails chaotic motion and destruction of the adiabatic invariants [24,25]. Only a finite region around the hyperbolic points is seed of chaotic motion, but the size of the region may be dramatically increased by allowing for time-dependent perturbations. For large enough perturbations, the position of the separatrices and of the X-points sweeps the whole available phase-space, thereby producing chaotic motions for practically any initial condition. The paradigm for this kind of systems is a charged particle in a uniform magnetic field interacting with a plane wave: in normalized units $H = \frac{p^2}{2} + \frac{x^2}{2} + A\sin(x - \omega t)$. Provided that the normalized wave amplitude $A$ exceeds unity, $H$ features a separatrix sweeping the phase space of area $O(A)$ around the origin. The existence of this mechanism does not rely upon details about the time behaviour of the wave, implying that even seemingly adiabatic scenarios (wave frequency ω much lesser than the cyclotron frequency, wave amplitude turned on and off infinitely slow) are not actually so, chaotic trajectories are actually irreversible. An important corollary is that, if the wave is turned on, a particle initially with very small kinetic energy is likely to acquire energy from the wave, which is no longer given back to the wave when it is switched off, as should occur in an adiabatic situation (see Fig. 6). This mechanism has been shown to work in astrophysical environments, since it quantitatively accounts for the observed temperatures of ions in the solar corona after interaction with Alfvén waves [26], but it is of interest in fusion scenarios as well, wherever charged particles interact with waves or with slowly changing conditions: actually, it was originally proposed as a heating possibility for electrons interacting with Lower-Hybrid waves in tokamaks [27].

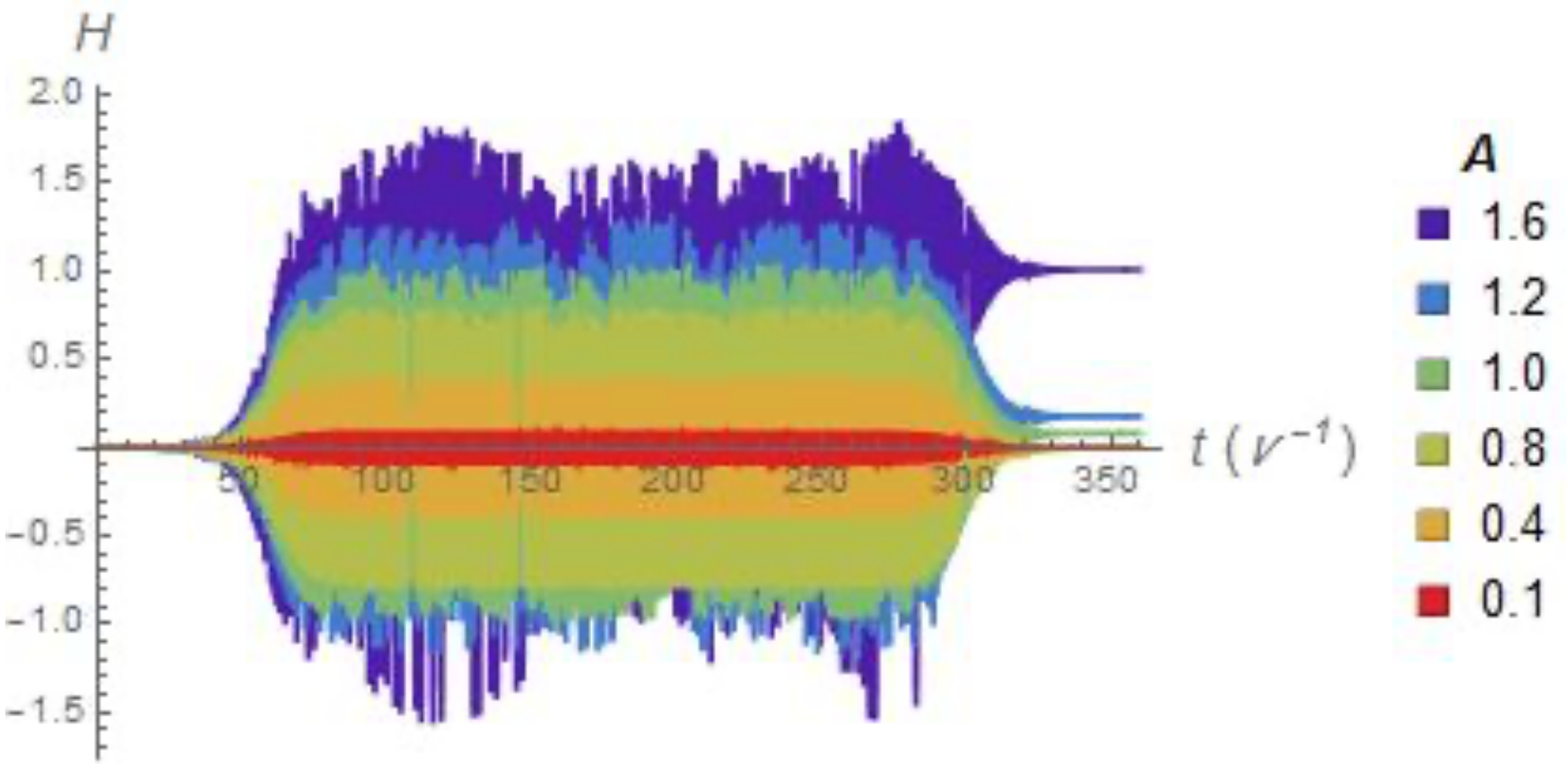

*FIG. 6. The figure shows time traces of particle' energy when the wave is slowly turned on and switched off. As long as the amplitude A is lower than unity, the dynamics is adiabatic and the particle return to its initial rest state. For A greater than unity, dynamics is non-adiabatic and the particle gains a net amount of energy.*

## 7. SUMMARY AND FINAL REMARKS

In this manuscript, most recent progress is reported concerning basic physics issues encountered in toroidal pinches. In particular, the current driven physics of helical self-organization (with barriers formation for magnetic field lines and temperature, temporary loss of operational point, reconnection-relaxation events) is addressed by relying on simple visco-resistive 3D nonlinear MHD modelling and related benchmarked numerical tools. A data analysis tool, machine learning "autoencoding" technique, is here trained for the first time on a RFP data analysis case. A possible mechanism for ion heating as produced by non-resonant low-frequency Alfvén waves is also presented, which was initially considered as possible mechanisms for RFP anomalous particle heating. Together with previous results on the discovery of new RFP helical states, experimentally reproduced in RFX-mod by suitable use of edge magnetic perturbations (MP), the recent progress give confidence in the realistic description obtained within our basic modelling, which might therefore provide a useful set of means to train and validate advanced data analysis tools, like machine learning techniques, with the aim of understanding and optimizing magnetic configurations.

## ACKNOWLEDGEMENTS

The authors acknowledge use of the computational resources provided by the EUROfusion High Performance Computer (Marconi-Fusion) through the EUROfusion project named 'PIXIE3D'.

## REFERENCES

[1] BONFIGLIO D., CHACÓN L., CAPPELLO S., Nonlinear three-dimensional verification of the SpeCyl and PIXIE3D magnetohydrodynamics codes for fusion plasmas, Phys. Plasmas **17** (2010) 082501.

[2] CIACCIO G., VERANDA M., BONFIGLIO D., CAPPELLO S., SPIZZO G., CHACÓN L., WHITE R.B., Numerical verification of ORBIT and NEMATO codes for magnetic topology diagnosis, Phys. Plasmas 20 (2013) 062505.

[3] PEGORARO et al, Coherent magnetic structures in self-organized plasmas, PPCF 61 (2019) 044003.

[4] CHACÓN L. et al., An asymptotic-preserving semi-Lagrangian algorithm for the time-dependent anisotropic heat transport equation, Jour. of Comput. Phys., 272, (2014) 719.

[5] BONFIGLIO D., VERANDA M., CAPPELLO S., ESCANDE D.F. AND CHACÓN L., Experimental-like helical self-organization in reversed-field pinch modelling Phys. Rev. Lett. 111 (2013) 085002.

[6] PIOVESAN et al., Role of a continuous MHD dynamo in the formation of 3D equilibria in fusion plasmas, Nucl. Fusion 57 (2017) 076014.

[7] VERANDA et al., Magnetohydrodynamics modelling successfully predicts new helical states in reversed-field pinch fusion plasmas, Nucl. Fusion 57 (2017) 116029.

[8] MARRELLI, et al., Upgrades of the RFX-mod reversed field pinch and expected scenario improvements, Nucl. Fusion 59, 076027 (2019) doi: 10.1088/1741-4326/ab1c6a.

[9] VALISA et al, High current regimes in RFX-mod, Plasma Physics and Controlled Fusion, 50 (2008) 124031.

[10] CAPPELLO S. et al, Bifurcation in Viscoresistive MHD: The Hartmann Number and the Reversed Field Pinch, Phys. Rev. Lett. 85, (2000) 3838.

[11] VERANDA M., et al, Helically self-organized pinches: dynamical regimes and magnetic chaos healing, Nucl. Fus. 60 (2020) 016007.

[12] PIOVESAN P, ZUIN M, et al. Magnetic order and confinement improvement in high-current regimes of RFX-mod with MHD feedback control. Nucl Fusion 49 (2009) 085036.

[13] SARFF, J. S., ALMAGRI, A. F., ANDERSON, J. K., et al, Overview of results from the MST reversed field pinch experiment, Nuclear Fusion (2013) 104017.

[14] DI GIANNATALE G, et al Submitted.

[15] DI GIANNATALE et al, Coherent transport structures in magnetized plasmas. 1, Phys. of Plas., 25(5) (2018) 052306.

[16] FALESSI, M., PEGORARO, F., & SCHEP, T., Lagrangian coherent structures and plasma transport processes, Journal of Plasma Physics 81(5), (2015) 495810505.

[17] LORENZINI R. et al, Self-organized helical equilibria as a new paradigm for ohmically heated fusion plasmas, Nat. Phys. 5 (2009) 570.

[18] MEHTA P., BUKOV M., WANG C.-H., DAY A.G.R., RICHARDSON C., FISHER C.K., SCHWAB D.J, A high-bias low-variance introduction to Machine Learning for physicists Physics Reports, 810 (2019) pp. 1-124.

[19] RIGONI GAROLA A., CAVAZZANA R., M. GOBBIN et al., Diagnostic data integration using deep neural networks for real-time plasma analysis (2020) arXiv:2010.15156 [physics.comp-ph] https://arxiv.org/abs/2010.15156.

[20] VERANDA, M., CAPPELLO, S., BONFIGLIO, D. et al. Magnetic reconnection in three-dimensional quasi-helical pinches. Rend. Fis. Acc. Lincei 31, (2020) 963–984.

[21] SPAGNOLO S., ZUIN M et al, Alfven eigenmodes in the RFX-mod reversed-field pinch plasma, Nuclear Fusion (2011) 083038.

[22] PFEFFERLÉ, D., GRAVES, J.P., COOPER, W.A., “Hybrid guiding-centre/full-orbit simulations in non-axisymmetric magnetic geometry exploiting general criterion for guiding-centre accuracy”, Plasma phys. Control. Fusion 57 (2015) 054017.

[23] GARREN, A., et al, Individual particle motion and the effect of scattering in an axially symmetric magnetic field, Proceedings of the second United Nations international conference on the peaceful uses of atomic energy, vol. 31, p. 65-70, United Nations, Geneva (1958).

[24] CAMBON, B., et al, Chaotic motion of charged particles in toroidal magnetic configurations, CHAOS 24 (2014) 033101.

[25] OGAWA, S., et al., Full particle orbit effects in regular and stochastic magnetic fields, Phys. Plasmas 23 (2016) 072506.

[26] ESCANDE, D.F., GONDRET, V., SATTIN, F., Relevant heating of the quiet solar corona by Alfvén waves: a result of adiabaticity breakdown, Sci. Rep. 9 (2019) 14274.

[27] DRAKE, J.F., LEE, T.T., Irreversibility and transport in the lower hybrid drift instability, Phys. Fluids 24 (1981) 1115-1125.